# The energy dependence of the electric dipole strength in heavy nuclei

Eckart Grosse[1,2], Frantisek Bečvář[3], Arnd R. Junghans[1], Gencho Rusev[1,*], Ronald Schwengner[1], Andreas Wagner[1]

[1]*Forschungszentrum Dresden-Rossendorf, Germany;*
[2]*Technische Universität Dresden, Germany;* [3]*Charles University, Prague, Czech Republic*

**Abstract:** On the basis of new photon scattering measurements and a reevaluation of average neutron resonance capture data we investigate how well Lorentzians adjusted to photo-neutron data in the giant dipole resonances give a good description of the photon strength also below the neutron threshold. If deformation effects are properly taken into account this is verified down to about 5 MeV for various nuclei with A>80 such that the previously employed differentiation between deformed and non-deformed nuclei is no longer necessary.

Similar to their mass and radius the response to electromagnetic radiation is a fundamental property of nuclei. Photo-nuclear processes were among the first nuclear reactions studied [1] and their appreciable strength has triggered the conclusion [2] that they are likely to play an important role for the nucleosynthesis: In the intense photon flux during high temperature cosmic scenarios particle emission thresholds are reached leading to the photo-disintegration of previously formed heavier nuclides. The photon strength is not only affecting the cosmic nucleosynthesis, it is of significance also for nuclear energy, as the photon strength and its energy dependence influence the analysis of radiative neutron capture data, of importance e.g. for the transmutation of radioactive waste. In contrast to the rather extensive experimental studies of the photon strength in the isovector giant dipole resonance (GDR) region well above the particle-separation energies, similarly detailed observations for the excitation region below are still surprisingly scarce and there exist inconsistencies in their description [3]. Apparently a standard two-resonance fit to GDR data in deformed nuclei works well to also describe the photon strength below particle thresholds, whereas in less deformed nuclei single Lorentzians seemed not to agree to low-energy data.

The work described here is motivated by the observation [4] that the extraction of electric dipole strength as applied in very many heavy nuclei is based on a fit of one Lorentz curve to the peak region of the GDR [5]. Using theoretical arguments [6], the resulting strength is then reduced at lower excitation energies where it seems to agree to some old primary photon data from neutron capture [7] assumed to deliver strength information at some energy below the particle thresholds. Various Hauser-Feshbach calculations use the resulting parameters as input and find deviations from new photo-dissociation data [8]. To derive a reliable parameterization of the photon strength we ignore the 'old' fits and instead look directly at the data in the GDR region. To avoid any sensitivity on particle parameters we only regard energies at which transmission coefficients are unity and data from below threshold are taken either from photon scattering or from average resonance neutron capture (ARC).

The photon strength function (PSF), $f_\lambda$ is related to the average photon absorption cross section

$$<\sigma_\gamma> = \vec{f}_\lambda(E_i=0, E_x) \cdot (2\lambda+1) \cdot (\pi\hbar c)^2 \cdot E_\gamma^{2\lambda-1}$$

and to the average gamma decay width following e.g. radiative capture [9]

$$<\Gamma_{E\lambda}> = \overleftarrow{f}_\lambda(E_i, E_f) \cdot D \cdot E_\gamma^{2\lambda+1}, \quad D = 1/\rho(E_i).$$

The Axel-Brink hypothesis states that the strength only depends on the transition energy and

detailed balance requires identity for absorption and emission processes:

$$\vec{f}_\lambda(E_1, E_2) = \overleftarrow{f}_\lambda(E_2, E_1) = f_\lambda(|E_1 - E_2|) = f_\lambda(E_\gamma).$$

The electric dipole strength is governed by the GDR resulting in a Lorentzian with a maximum at $E_k$:

$$f_1(E_\gamma) = \frac{2 \cdot I}{3\pi(\pi\hbar c)^2} \frac{E_\gamma \cdot \Gamma_0}{\left(E_0^2 - E_\gamma^2\right)^2 + E_\gamma^2 \Gamma_0^2}.$$

The resonance integral $I = \int \sigma_\gamma \cdot dE_\gamma$ is the appropriate measure of the size of the resonant cross section. The GDR centroid energies $E_0$ of a spherical nucleus with A and Z are well predicted by the FRDM [10] with the symmetry-energy constant J = 32.7 MeV and the surface stiffness Q = 32.7 MeV taken from the fit to masses. The additional parameter $m_{eff} \cdot c^2$ = 874 MeV gives the best fit to all nuclei with A>80. Traditionally GDR strength and width were extracted from a fit of a Lorentzian to (γ,n) cross sections near the maximum [5, 11].

This procedure appears as straightforward, but it has disadvantages:
1. Fit parameters for width and strength are strongly correlated.
2. Data for channels other than (γ,n) are not properly included.
3. (γ,n)-data suffer from uncertainties; a later yield correction [11] is not included.
4. The Thomas-Reiche-Kuhn (TRK) sum rule is very often not fulfilled.
5. No account is made for triaxiality well known from nuclear spectroscopy.

Nuclear deformation results in a split of the GDR; for nuclei not identified as deformed this misleadingly indicates an increase in spreading width by up to 3 MeV.

We propose a different approach to parameterize the dipole strength in the GDR and especially account for deformation effects in a direct way by using independent information from spectroscopic studies on the deformation parameters β and γ. As the GDR dipole oscillation is fast as compared to vibrational or rotational modes the widening due to the deformation can be treated adiabatically by inserting mean deformation parameters in the Hill-Wheeler expression:

$$E_k = E_0 \cdot \exp[-\sqrt{5/4\pi} \cdot \beta \cdot \cos(\gamma - \tfrac{2}{3} k\pi)] \, .$$

Generalizing a suggestion of Bush & Alhassid [12], originally formulated for three components of one GDR belonging to the three orthogonal nuclear axes, we parameterize the spreading width for all A>80 by $\Gamma_k(E_k) = 1.99$ MeV $\cdot (E_k/10\text{MeV})^{1.6}$ with the exponent 1.6 derived from hydrodynamical considerations and the proportionality factor obtained from a fit to more than 20 different nuclei with A>80. Distributing the integrated dipole strength $I_{TRK}$ as predicted by the TRK sum rule evenly over the three components we obtain the expression:

$$f_1(E) = \sum_{k=1,3} \frac{2 \cdot I_k}{3\pi(\pi\hbar c)^2} \frac{E_\gamma \cdot \Gamma_k}{\left(E_k^2 - E_\gamma^2\right)^2 + E_\gamma^2 \Gamma_k^2}, \quad I_1 = I_2 = I_3 = \tfrac{1}{3} I_{TRK}$$

which we compare to data
1. to determine their deviation from the TRK sum rule and
2. to find out to what extent it also describes the dipole strength at low energies.

To cover a wide range in A and in deformation we present here results for $^{88}$Sr, $^{98}$Mo, $^{148}$Sm, $^{168}$Er, $^{190}$Os and $^{197}$Au. For these nuclei reliable data in the GDR region [13] and below the neutron threshold are available. If for the latter ARC-data are used, their cross section had to be normalized such that strength below the gamma ray detection limit is accounted for [14, 15]. In the photon scattering studies at ELBE [16] this was achieved by adding up all nuclear scattering yield observed in large volume Ge-detectors combined to Compton-escape suppression-shields.

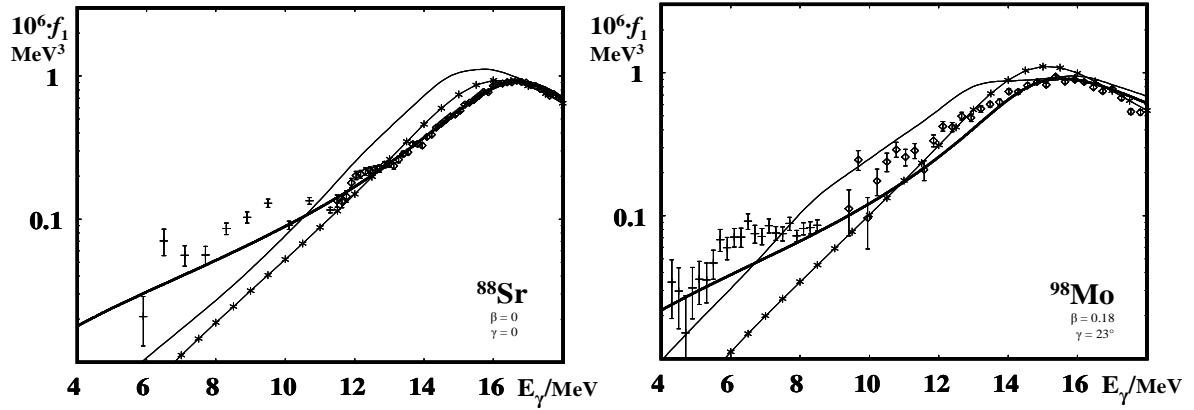

**Fig.1:** Dipole strength below and above the neutron threshold for $^{88}$Sr and $^{98}$Mo.
The experimental data from photon scattering [+, 16] and ($\gamma$,n) [◊, 17] are averaged over 0.25 (resp. 0.6) MeV to reduce the influence of Porter-Thomas fluctuations.
The thick line represents our parameterization whereas the thin line corresponds to E1-strength functions as determined within the QRPA model based on the SLy4 Skyrme force as presented in RIPL-2 [18].
The analytic expression 'EGLO' as tabulated in RIPL-2 [18] is shown as —∗—.

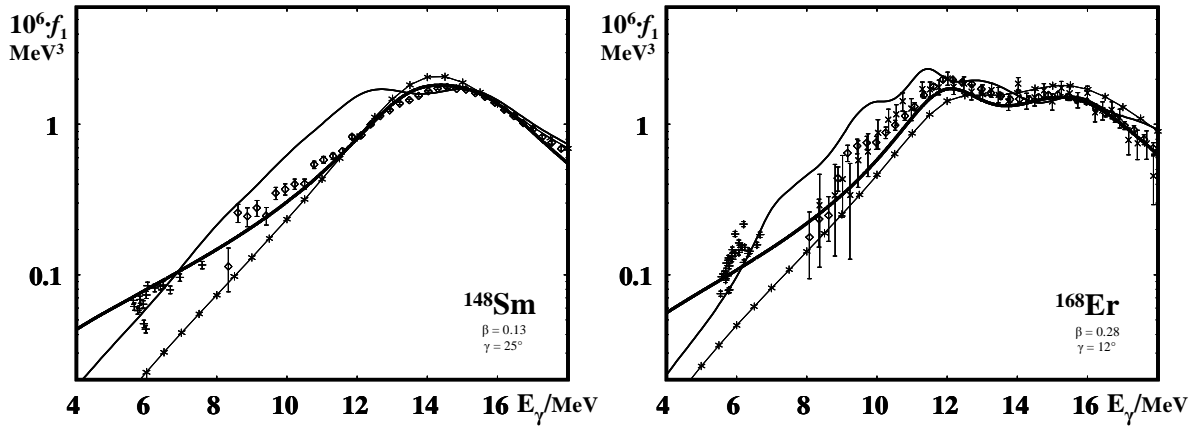

**Fig.2:** Dipole strength below and above the neutron threshold in the nuclei $^{148}$Sm and $^{168}$Er.
The experimental data from ARC [+, 15], ($\gamma$,n) [◊, 19] and photon absorption [x, 20] are averaged over 0.2 MeV. The lines and symbols without error bars have the same meaning as in the top figures.

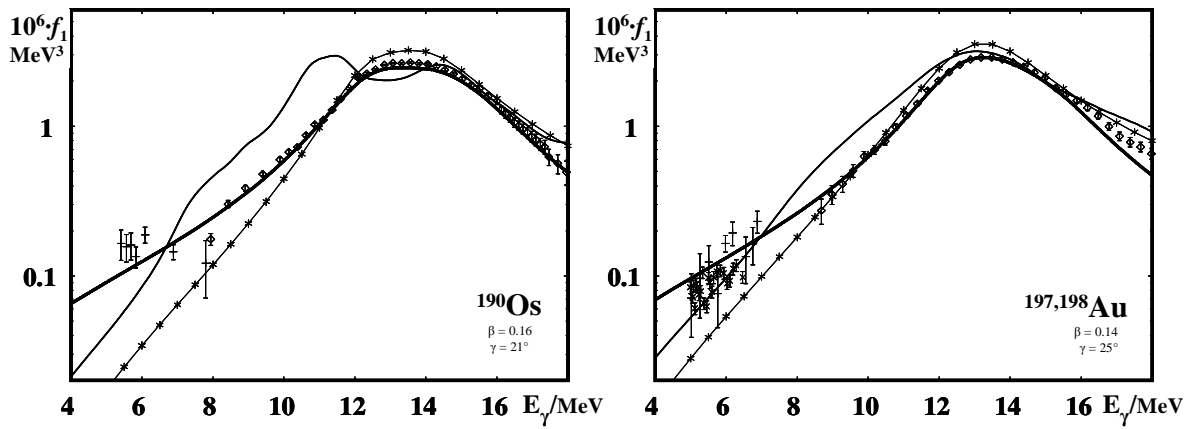

**Fig.3:** Dipole strength in the nuclei $^{190}$Os and $^{197}$Au: Data for above the neutron threshold [19, 21] are combined to experimental data from photon scattering [+, 9] and ARC [x, 14, 15]. They agree to each other within their error bars. The lines have the same meaning as in the top figures.

Obviously the two photon strength functions taken from RIPL-2 [18] deviate from the data especially below the neutron thresholds. The other of the PSFs proposed there do not much better and also the Hauser-Feshbach type calculations [22,23] often quoted [8] may have deficiencies due to the fact that they also rely on the 'old' analysis [5] of ($\gamma$,n) data. From the favorable comparison of the parameterization as proposed by us to the various experimental data the following conclusions can be drawn:

1. for all A>80 and $E_\gamma$ >5 MeV a fit to GDR's requires only 2 free parameters :
   a. $E_0$ is predicted by FRDM with m*·c² = 874 MeV as additional parameter.
   b. The spreading can be parameterized by
      $\Gamma_k = \Gamma_0 \cdot (E_k/E_0)^{1.6}$ with $\Gamma_0$ ($E_0$=10 MeV) = 1.99 MeV.
2. The photon strength down to 5 MeV is described well, if spectroscopic information on deformations β and γ is introduced; no $E_\gamma$-dependence of Γ is needed for $E_\gamma$ > 4 MeV – at variance e.g. to the KMF-ansatz [6] based on theoretical arguments.
3. The dipole strength in near magic and in well deformed nuclei can be described with the same formalism – in contrast to statements [3] made previously.
   For transitional and triaxial nuclei our parameterization works equally well, if spectroscopic information on their deformation is properly included.
4. The strength in the observed range deviates by less than 9 % from the TRK sum rule with (pygmy) strength fragments below the GDR contributing a few percent only.
5. As indicated for $^{197}$Au the strength data below $S_n$ as obtained from photon scattering [9] and from gamma emission after ARC [15] agree to each other – in accordance to the Axel-Brink hypothesis.
6. There is an obvious need for such a comparison in other nuclei as well.

It should be noted here that radiative neutron capture data do not allow clear conclusions for photon energies below 4 MeV. From photon scattering one knows that at these low energies various discrete levels can be excited by E1, E2 or M1 photons. For magnetic excitations predictions [24] have been made combining calculations for spin-flip and orbital modes; electrically excited modes strongly depend on nuclear deformation and softness and cannot easily be described by a simple parameterization.

Concerning the use of photon strength parameterizations in Hauser-Feshbach calculations for nuclear astrophysics or for nuclear energy applications the parameterization as presented here seems to be superior to the various prescriptions proposed previously e.g. by RIPL-2 [18]. As these strength functions – as well as variations given in the codes TALYS [22] and NON-SMOKER [23] are based on the 'traditional' extraction of GDR parameters [5] they should be used with care – as we have pointed out the shortcoming of this extraction method.

We are presently exploring if our parameterization can be made more predictive by combining it to prescriptions relating the quadrupole deformation to the number of valence nucleons [25]; for the triaxiality γ a relation to β has been established [26] for the mass range as discussed here. Concerning possible corrections to our method necessary to account for the distribution of deformation values instead of our use of average numbers the respective theoretical work ongoing at FZD should be mentioned – it was presented [27] at CSG-13.

Concluding: The giant dipole mode has very much the same properties in all heavy nuclei when their deformation is properly accounted for and its spreading only varies (very smoothly) with the resonance energies $E_k$ and <u>not</u> with the energy $E_\gamma$ of the absorbed photon.

**\*)** Present address: Physics Department, Duke University, Durham NC, USA.